\documentclass[journal]{IEEEtran}
\usepackage[utf8]{inputenc}
\usepackage{graphicx}
\usepackage{amssymb}
\usepackage{amsmath}
\usepackage{makecell}
\usepackage{multicol}
\usepackage{authblk}
\usepackage{float}
\usepackage{multirow,tabularx}
\usepackage{cite}
\usepackage[dvipsnames]{xcolor}
\begin{document}

\title{Stochastic Resource Allocation for Semantic Communication-aided Virtual Transportation Networks in the Metaverse}

\author{Wei Chong Ng$^{1,2}$, Hongyang Du$^{3,4}$, Wei Yang Bryan Lim$^{2}$, Zehui Xiong$^{5}$, Dusit Niyato$^{4}$ and Chunyan Miao$^{4,6}$\\
\vspace*{-4mm}$^1$Alibaba Group~$^2$Alibaba-NTU Joint Research Institute~
$^3$Energy Research Institute @ NTU\\
$^4$School of Computer Science and Engineering, Nanyang Technological University, Singapore \\
$^5$Singapore University of Technology and Design\\ 
\vspace*{-10mm}}
\maketitle

\begin{abstract}
The physical-virtual world synchronization to develop the Metaverse will require a massive transmission and exchange of data. In this paper, we introduce semantic communication for the development of virtual transportation networks in the Metaverse. Leveraging the perception capabilities of edge devices, virtual service providers (VSPs) can subscribe to their preferred edge devices to receive the semantic data of interest. However, the demands of the VSPs are highly dependent on the users that they are serving. To address the resource allocation problem amid stochastic user demand, we propose a stochastic semantic transmission scheme (SSTS) based on two-stage stochastic integer programming. Using real data captured by edge devices we deploy in Singapore, the simulation results show that SSTS can minimize the transmission cost of the VSPs while accounting for the users' demand uncertainties.
\end{abstract}

\section{Introduction}
The concept of the Metaverse first appeared in 1992~\cite{stephenson2003snow}. However, it was in  recent years that it received much attention from academia and industry due to the growing feasibility of realizing the Metaverse as a result of technological advancements, e.g., mixed/augmented/extended/virtual reality (MR/AR/ER/VR), 6G network and edge computing. For example, Meta spent $\$10$ billion to start building the Metaverse in 2021. With the recently developed Horizon Worlds VR social platform, users can trade virtual items and be encouraged to create their own contents~\cite{metainvest}. 

The Metaverse can be seen as an integration of multiple virtual worlds accessible using technologies such as VR/AR and developed using artificial intelligence (AI)~\cite{chu2022metaslicing}. Besides, a defining characteristic of the Metaverse is the closely-linked physical and virtual domains. On one hand, virtual worlds within the Metaverse can be constructed using transmitted/perceived data from edge devices such as the Internet of Things (IoT). On the other hand, the physical domain is influenced by actions taken in the virtual world, e.g., through IoT actuation.

To realize the Metaverse, it is therefore crucial that future communication systems are capable of supporting the transmission and exchange of tremendous amounts of data. However, the technical requirements for constructing high-resolution virtual worlds that accurately reflect the physical environment in a timely manner are more stringent than what current fifth-generation (5G) networks may support~\cite{luo2022semantic}. In response, semantic communication systems may be instrumental for the development of the Metaverse. Unlike existing communication technologies, the transmission in semantic communication is considered effective as long as the received information retains the same meaning as the transmitted information. For example, when a user requires an image, semantic communication systems utilize \textit{semantic extraction} to reduce the transmitted data such that only the region in which the user is interested in is transmitted.

In this paper, we propose a case study of developing a virtual transportation network in the Metaverse using real data~\cite{bigdata}. We refer to a VSP as an entity that provides a virtual service in the Metaverse. Using data from the physical domain such as weather conditions or images of geographical landmarks and vehicles, the VSP is able to provide immersive experiences to users, e.g., for realistic test driving of vehicles or the safe training of autonomous vehicles subjected to practical constraints. The captured data from physical domains may usually be traded in data markets  or retrieved through crowdsensing \cite{niyato2016market}. Specifically, edge devices may sell data captured from geographical regions in which they are deployed. 

Subscription plans are required so that the edge devices are paid for when the VSPs use semantic data transmission. Following the data pricing model of~\cite{ng2022unified}, there are two types of subscription plans in general: reservation and on-demand. The reservation plan allows the VSPs to purchase the number of transmissions in bundles while the on-demand plan charges the VSP per transmission. The on-demand plan is a one-time, short-term plan, so it is more expensive than the reservation plan. However, with the uncertainty in demands, it is difficult for VSPs to choose the optimal subscription strategy. For example, the detection system may not be able to detect pedestrians very well under extreme weather conditions such as thunderstorms. Hence, the on-demand plan may be triggered to obtain more data to update the machine learning model. Moreover, the VSPs may have different interests in the types of semantic data (images) required. For example, an autonomous vehicle VSP may require more semantic data from edge devices deployed around a particular region. To minimize the operation cost of the VSP while addressing the demand uncertainty of the users, we propose a two-stage stochastic integer programming (SIP) scheme for semantic data subscription provisioning. The contributions of this paper are summarized as follows.

\begin{itemize}
    \item We, for the first time, introduce the integration of semantic communication and stochastic integer programming to devise fully dynamic transmission solutions for emerging AI-enabled Metaverse applications in the virtual transportation network.
    \item Our proposed stochastic semantic transmission scheme (SSTS) minimizes transmission/storage costs and energy costs within the network given the uncertainty of users’ demands. The scheme is capable of handling the unknown by incorporating recourse actions to remedy the under-subscription incident.
    \item Using the real data, we demonstrate that SSTS achieves superior performances compared with other baselines such as Expected-Value Formulation (EVF) and random allocation schemes.
\end{itemize}

The remainder of the paper is organized as follows: In Section~\ref{system_model}, we present the network model. In Section~\ref{problem}, we formulate the problem. We discuss and analyze the simulation result in Section~\ref{simulation}. Section~\ref{conclusion} concludes the paper.

\begin{figure}[htp]
    \centering
    \includegraphics[width=7cm, height=4cm,trim={0cm 20cm 5cm 0cm},clip]{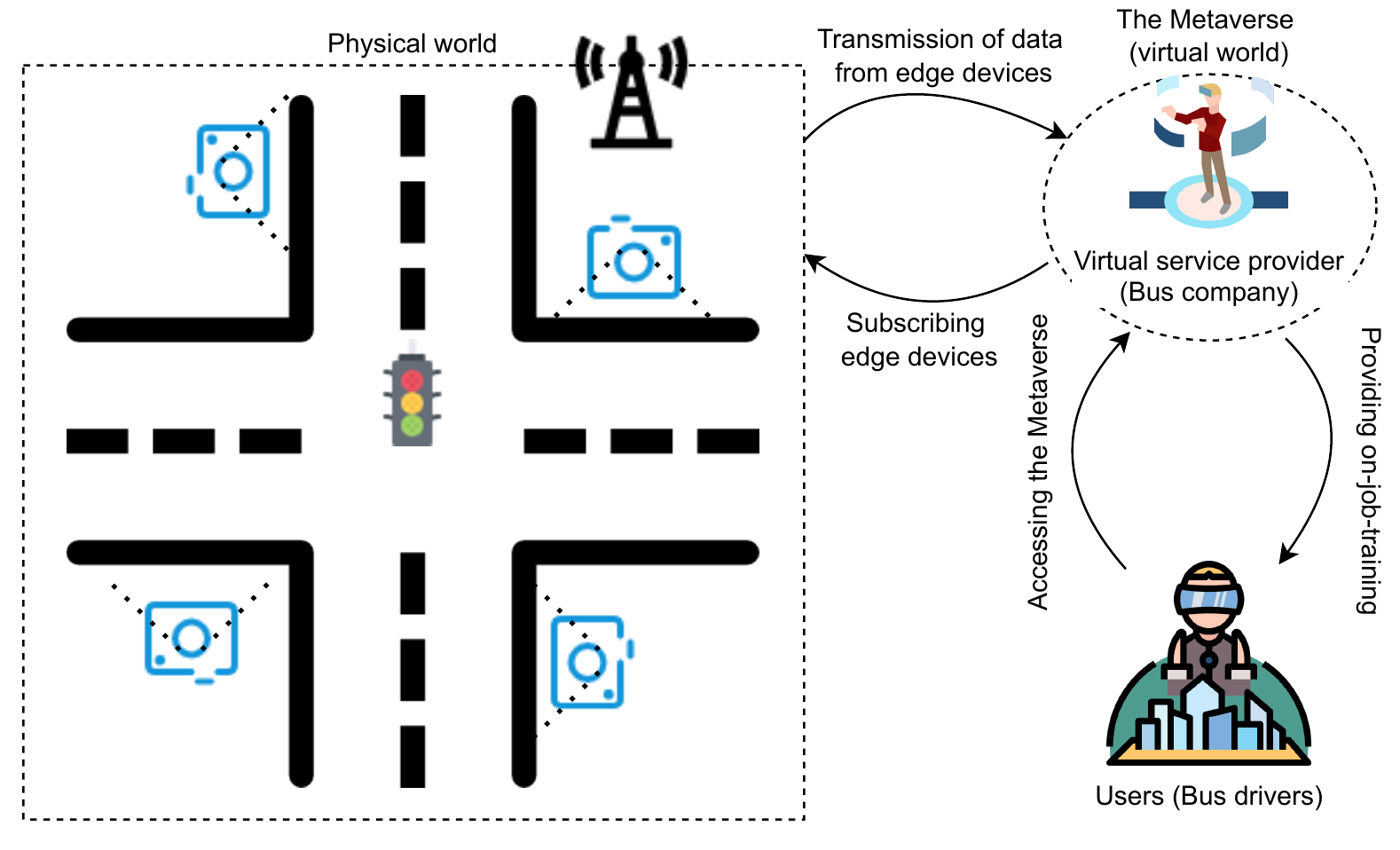}
    \caption{An illustrative example of the system model with one VSP from a bus company and four edge devices and they are placed at the traffic junction. The VSP is from an autonomous company and has an interest }
    \label{fig:network}
\end{figure}

\section{System Model}\label{system_model}
We consider the system model from the perspective of VSPs participating in the Metaverse. Let $\mathcal{W} = \{1,\ldots,w,\ldots,W\}$ denote the set of VSPs and $\mathcal{E}~=\{1,\ldots,e,\ldots,E\}$ denote the set of edge devices with sensing capabilities in the physical world. Each edge device has a different field of view pointed to different locations. Figure ~\ref{fig:network} shows an illustrative example of the system model with a VSP that operates a bus company. The VSP uses the Metaverse as a platform to provide on-the-job training to bus drivers (users). The four edge devices are situated at the traffic junction to capture images (semantic data) that are related to vehicles as most of the time, traffic accident happens at the junction~\cite{bruede1999accidents}. The edge devices such as smartphones can transmit semantic data to the respective Metaverse VSPs through a base station (BS). Using the received data, VSPs can provide a more realistic Metaverse application to the users. 

\subsection{The Edge Data Pricing Model}
Before the data are transmitted to the VSPs, the VSPs have to subscribe to the resources  in advance to secure them as separate entities (edge devices) own these resources. There are two types of subscription plans, i.e., reservation and on-demand plans. The reservation plan is treated as the long-term plan, which allows the VSPs to maintain a long-term collaboration with the edge devices. In contrast, the on-demand plan is a short-term plan. It is used only when the edge device's service is needed temporarily, e.g., when the reservation plan is insufficient to meet the semantic data demand. 

To enjoy the reservation plan, the VSPs have to pay a membership fee (monthly) to the owner of the edge devices. This entitles the VSP to a lower costing bundle. Note that each bundle supports $n$ transmissions for the VSP, i.e., one transmission includes that the edge device performs sensor data collection, semantic extraction, and wired or wireless transfer. Let $C_e^{(r,memb)}$ denote the membership cost of each edge device and $C_e^{(r,trans)}$ represent the reservation cost for each bundle charged by edge device $e$. However, each VSP that is providing services in the Metaverse has a different demand from its users, and the demand is not known when the reservation of the edge devices is made. Consider the situation that, autonomous vehicles (AVs) are expensive and unsafe to train on the physical road involving other vehicles and  pedestrians. Instead, the virtual world can be a good medium to train AVs in realistic scenarios. For example, the company Oxbotica uses the Metaverse to improve its AVs' object detection algorithms~\cite{Oxbo}. As an illustration, the VSPs serving the users may request 100 semantic road data to train the AV further. However, if the weather changes rapidly, the VSP will require more semantic data that corresponds to the unforeseen weather scenario, e.g., 400 transmissions instead. Therefore, in this case, an on-demand plan is needed to accommodate the shortfall. The cost of each on-demand transmission is denoted as $C_e^{(o,trans)}$.

\subsection{Uncertainty in Demands}
With the uncertainty mentioned above, the demand for VSPs is not always fixed. Let $\lambda_i$ denote the  demand scenario $i$ of all the VSPs. The set of scenarios is denoted by $\Omega$, i.e., $\lambda_i\in\Omega$. Let $P(\lambda_i)$ denote the probability that scenario $\lambda_i\in\Omega$ is realized, where $P(\lambda_i)$ can be obtained from the historical records~\cite{ng2022unified}. The uncertainty of demands is expressed as follows:

\begin{center}
{\footnotesize
$\lambda_i =\begin{bmatrix}
\arraycolsep=3pt
\medmuskip=1mu 
(F_{1},\bar{F}_{1},\Tilde{F}_1),\ldots,(F_{W},\bar{F}_{W},\Tilde{F}_W)
\end{bmatrix}$,}
\end{center}

\noindent where $F_{W}$ represents the interest of VSP $W$, $\bar{F}_{W}$ represents the number of semantic data transmissions that VSP $W$ requires, and $\Tilde{F}_W~\in~\{0,1\}$ represents the threshold of VSP $W$. For example, $\lambda_i=~\{(F_{W}~:~\text{traffic conditions},~\bar{F}_{W}~:~100~,~\Tilde{F}_W~:~0.8)\}$ means that VSP $W$ has an interest in traffic condition data. Also, out of the 100 semantic data required, it is acceptable when the edge device can only provide 80 semantic data relevant to the VSPs' interest, as the VSPs can achieve their objective by using 80 semantic data.


\subsection{Transmission Model}
We assume that each edge device is allocated with an orthogonal spectrum resource block to avoid the co-interference among the edge devices~\cite{zhou2019energy}. Let $r_{e}$ denote the uplink data transmission rate from edge device $e$ to the BS under its coverage. Then, the transmission time taken $t_e$ is defined as follows~\cite{fan2017computation}:
\begin{align}
    t_e = \frac{q_e}{r_{e}},
\end{align}
where $q_e$ is the size of the transmitted data. Let $w$ denote the transmit power used by edge device $e$ in the uplink transmission, and the energy consumption of the transmission is represented as follows~\cite{fan2017computation}:
\begin{align}
    x_{e} = wt_{e}.
\end{align}
Therefore, using the energy consumption model, $C_e^{(r,trans)}~=~\frac{\hat{n}w\bar{q}}{r_e}\alpha_1$ and $C_e^{(o,trans)}=\frac{w\bar{q}}{r_e}\alpha_2$. For simplicity, the average transmitted data size $\bar{q}$ is used, and it is obtained from past historical records. $\hat{n}$ is the number of semantic data transmissions. $\alpha_1$ and $\alpha_2$ are the cost coefficient associated with energy consumption, where $\alpha_2>\alpha_1$ since the on-demand plan is typically more expensive than the reservation plan.

\subsection{Category Generation and Similarity Matching}\label{sentencegeneragtion}
Using the plans, the VSPs can obtain semantic data captured by the edge devices. However,  it is difficult to identify which edge device produces images that are important or relevant to the interests of the VSPs. This paper adopts a pre-trained machine learning model, you only look once (YOLO) from~\cite{redmon2018yolov3}. With the help of YOLO, objects (semantic data) and their corresponding categories can be extracted from the respective images. YOLO provides real-time detection with relatively high accuracy. When the category is within the interest of the VSPs, the semantic data, which is the snapshot of the object, can be transmitted.

Once the categories are generated from the images, the VSPs cannot identify which edge device to subscribe to as the VSPs do not know which semantic data is relevant to their interest. It is not practical for the VSPs to search manually through the categories. Therefore, we propose to use category similarity for the VSPs to subscribe to the edge device that produces the best semantic data that meets the interest of the VSPs. However, in different contexts, the same word might have multiple definitions. For example, ``wind'' can mean the current of air or the action turn. The traditional method, such as \textit{word2vec} cannot recognise polysemy. The issue arises when the same word cannot be represented by the same numerical vector in different contexts. One of the solutions is to use BERT~\cite{xie2021deep}. BERT is a powerful pre-trained machine learning model that has been trained by billions of sentences for extracting semantic information. It is used to convert multiple categories and convert them into vectors according to different contexts. In this paper, we use cosine similarity~\cite{xie2021deep} to measure the similarity between the two sentences. Specifically, 
\begin{align}\label{similar}
    \text{match(\textbf{A},\textbf{B})} = \frac{\textbf{A}\cdot\textbf{B}}{\|\textbf{A}\|\|\textbf{B}\|},
\end{align}
where \textbf{A} is the vectorized of $F_w$ and \textbf{B} is the vectorized output generated from BERT. The category similarity defined in(~\ref{similar}) is a number between 0 and 1, which indicates how similar \textbf{A} is to \textbf{B}, with 1 representing the highest similarity and 0 representing no similarity. The average similarity of VSP $w$'s demands and edge device $e$'s data types is represented by $S_{w,e}$ and can be calculated from $S_{w,e}=\frac{\sum_{i=1}^y \text{match}(\textbf{A},\textbf{B}_i)}{y}$, where $y$ is the total number of images in edge device $e$. As a result, the VSPs can subscribe to receive the semantic data from the edge device that has the highest similarity score. 
%

\begin{figure}[htp]
    \centering
    \includegraphics[width=8.5cm, height=10cm,trim={0.2cm 9.4cm 0.2cm 0cm},clip]{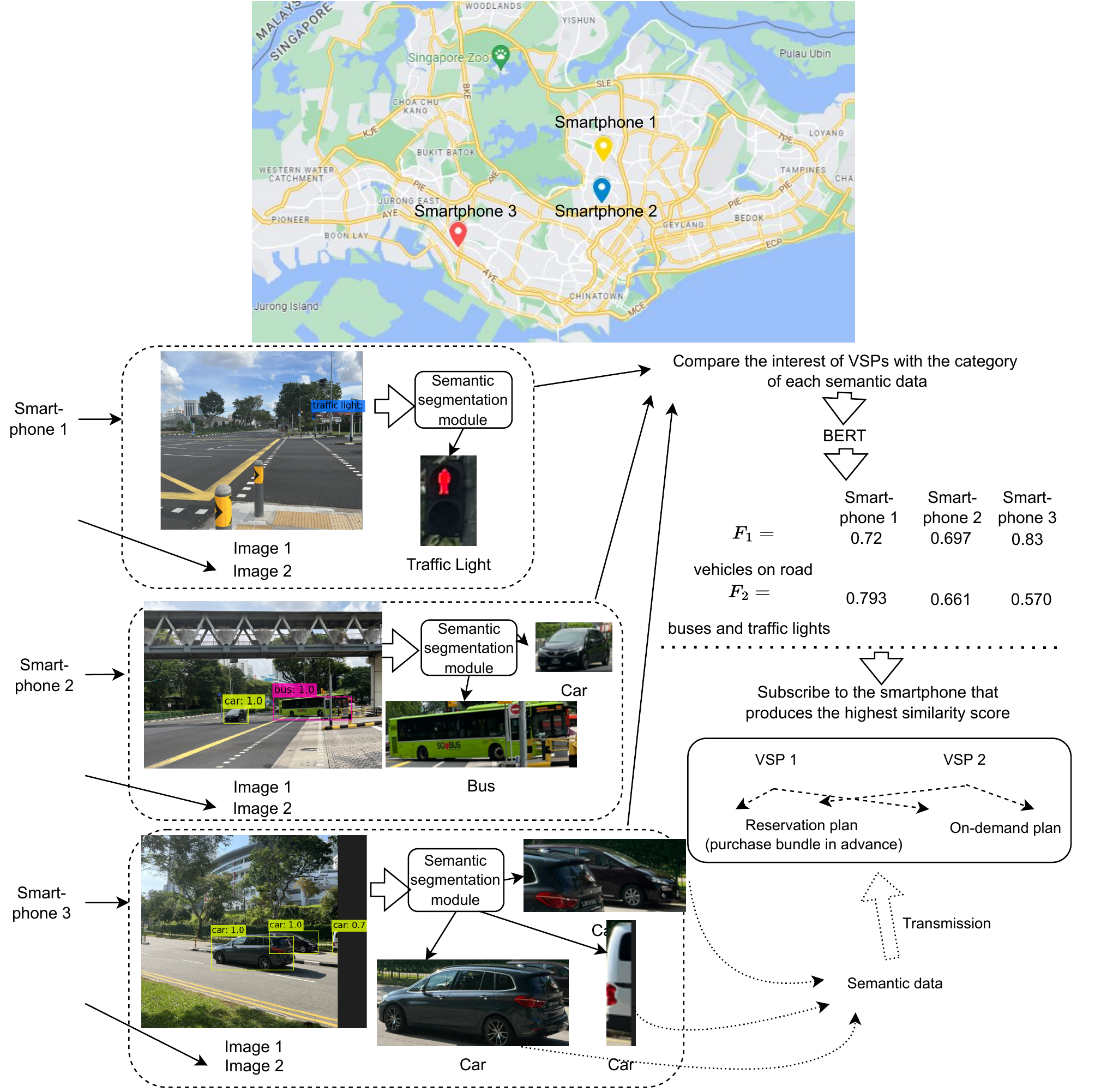}
    \caption{Locations of the smartphones in Singapore and an illustrative example of a virtual transportation case study.}
    \label{fig:case}
\end{figure}

\subsection{Virtual Transportation Case Study}\label{case_study}

A virtual transportation network is used as a case study to illustrate the system model. In this example, there are two VSPs, VSP 1 and VSP 2. VSP 1 is a company to provide the service of the AV, while VSP 2 is a bus company. AVs are dangerous to travel on the road when the machine learning model is not fully trained. However, the vehicles are unable to train when they are not allowed to travel on the road. Therefore, one solution for VSP 1 is to set up a simulated environment in the Metaverse, which digitises the physical road using real data as input. VSP 1 can then test the AVs' detection systems in the  Metaverse and update the machine learning model depending on users' demands in the physical world.

Similarly, it is also dangerous for a pedestrian when VSP 2 conducts their on-the-job training for new bus drivers as the new drivers are unfamiliar with the traffic conditions or inexperienced. Therefore, virtual on-the-job training can be conducted in the Metaverse. VSP 2 can replicate the transportation networks in the physical world into the Metaverse. However, in order to support the Metaverse, VSPs 1 and 2 require a tremendous amount of data from the edge devices in the physical world to support their Metaverse. For example, VSP 1 may have the interest of $F_1=$ vehicles on the road. Different vehicle images can be used to constantly update the database to improve the Metaverse for autonomous vehicles. Meanwhile, VSP 2 may have the interest of $F_2=$ images related to the buses and traffic lights so that VSP 2 can monitor the physical driving condition of the bus driver and improve their virtual on-the-job training procedure.


In order to provide a more realistic or practical scenario, we deploy three different smartphones to act as edge devices at three different locations in Singapore. The locations are shown in Fig~\ref{fig:case}. The VSPs do not know whether the images captured by smartphones are related to their interest. Therefore, before the VSPs subscribe to which smartphone to use, each smartphone owner provides $\bar{n}$ data for the VSPs to study the historical records. From the historical captured images, the VSPs can extract the category and also the corresponding snapshots (semantic data). An average similarity score can be obtained from the historical category generated by the respective smartphones used to compare with the interest of the VSPs. Figure~\ref{fig:case} is used to illustrate the case study. The average similarity score helps the VSPs choose the optimal smartphone to subscribe, for example, using the average similarity score obtained in Fig~\ref{fig:case}. When the threshold is 100\% $\Tilde{F}_w=1$, $n=120$ and the number of semantic data transmissions required is $\bar{F}_w=100$, it is definitely cheaper for the VSP 1 to subscribe smartphone 3. As out of the data transmitted, 83\% of the data is related to the interest of VSP 1. Smartphone 3 can meet the demand of VSP 1 by using only 1 bundle, $1\times n\times83\%\geq\bar{F}_w\times\Tilde{F}_w$. The snapshots of the objects (semantic data) are transmitted to the respective VSPs once the smartphone is subscribed. However, each VSP experiences different user demands, making it difficult for VSPs to subscribe for the optimal plan. Therefore, this paper uses a stochastic allocation scheme to optimize the resources by considering the demand uncertainties.

\section{Problem Formulation}\label{problem}
This section introduces Deterministic Integer Programming (DIP) and Stochastic Integer Programming (SIP) to optimize the resources used by minimizing the total cost of the VSPs.

\subsection{Deterministic Integer Programming}\label{dip}
The VSPs can subscribe to the optimal edge devices and purchase the optimal number of bundles for semantic data transmission by using the reservation plan when the demand is precisely known. Therefore, the on-demand plan is not required. In total, there are two decision variables.
\begin{itemize}
    \item $m_{w,e}^{(\mathrm{r})}\in\{0,1\}$ indicates whether VSP $w$ pays the membership cost to the owner of the edge device $e$.
    \item $\bar{m}_{w,e}^{(\mathrm{r})}\in\{0,1,\ldots\}$ indicates the number of bundles purchased by the VSP $w$. 
\end{itemize}

A DIP can be formulated to minimize the total cost of the VSPs as follows:

\noindent $\displaystyle\min_{m_{w,e}^{(\mathrm{r})}, \bar{m}_{w,e}^{(\mathrm{r})}}$:

\begin{align}\label{dip1}
    \sum_{w\in \mathcal{W}}\sum_{e\in\mathcal{E}}\biggl(m_{w,e}^{(\mathrm{r})}C_e^{(r,memb)} + \bar{m}_{w,e}^{(\mathrm{r})}C_e^{(r,trans)} \biggr),
\end{align}
subject to: 
\begin{align}
    &\bar{m}_{w,e}^{(\mathrm{r})}\leq m_{w,e}^{(\mathrm{r})}X, &\forall w\in\mathcal{W}, \forall e\in\mathcal{E},\label{dip_cons1}\\
    &\sum_{e\in\mathcal{E}}\bar{m}_{w,e}^{(\mathrm{r})}nD_{w,e} \geq \bar{D}_{w}\Tilde{D}_{w}, &\forall w\in \mathcal{W},\label{dip_cons2}\\
    & m_{w,e}^{(\mathrm{r})}\in\{0,1\}, &\forall w\in\mathcal{W}, \forall e\in\mathcal{E},\label{dip_cons3}\\
    & \bar{m}_{w,e}^{(\mathrm{r})}\in \mathbb{Z}\text{-}*, &\forall w\in\mathcal{W}, \forall e\in\mathcal{E}.\label{dip_cons4}
\end{align}

The objective function in~(\ref{dip1}) is to minimize the total cost due to subscription reservations. $D_{w,e}$ is the actual average similarity score between the interest of the VSP $w$ and the edge device $e$. $\bar{D}_{w}$ is the actual image demand of VSP $w$ and $\Tilde{D}_w$ is the actual acceptable threshold of VSP $w$. The constraint in~(\ref{dip_cons1}) ensures that the VSP has to pay the membership fee to the edge device owner before the VSP can purchase any bundle from the respective edge device. (\ref{dip_cons2}) ensures that the demand is met. For example, when the number of edge devices is 1, $e=1$, $n=200, D_{w,1}=0.8$, $\bar{D}_w=200$, and $\Tilde{D}_w=1$. It means that only $80\%$ of the data that are captured by the edge device $1$ is relevant to the interest of VSP $w$. VSP $w$ faces a shortfall of $20\%$, and the acceptable threshold of $w$ is 100\%. As a result, instead of 1 bundle, VSP $w$ has to purchase two bundles $m_{w,1}^{(\mathrm{r})}=2$ to meet the demand of 200 semantic data transmissions. (\ref{dip_cons3}) indicates that $m_{w,e}^{(\mathrm{r})}$ is a binary variable. (\ref{dip_cons4}) indicates that $\bar{m}_{w,e}^{(\mathrm{r})}$ is a non-negative integer.

\subsection{Stochastic Integer Programming}\label{sip_pro}
If the demands of the VSPs are not known, the DIP formulated in~(\ref{dip1})~-~(\ref{dip_cons4}) is no longer applicable. Therefore, SIP with a two-stage recourse is developed. This section introduces the SIP to minimize the total cost of the network by optimizing the number of edge devices to subscribe to and the number of semantic data transmissions to subscribe. The first stage consists of all decisions that must be selected before the demands are realized and observed. The VSPs have to subscribe to the edge device and purchase the corresponding bundle before observing the demands. In the second stage, decisions are allowed to adapt to the demand observed. After the demand is observed, the VSPs have to pay for the additional images needed if the number of images reserved in the first stage is lesser than the demand.

\begin{table}
\centering
\caption{Experiment parameters}
\begin{tabular}{|l| c|} 
 \hline
  \textbf{Parameter} & \textbf{Values} \\ 
 \hline
 $\alpha_1$                                              & 5\\
 $\alpha_2$                                             & 15\\
 Membership cost~\cite{cameracost}, $C_e^{r,memb}$                       & $\$1.89$  \\ 
 Uplink data transmission rate~\cite{fan2017computation}, $r_e$                 & $[1.5,3.5]$ MB/s  \\ 
 Uplink transmission power~\cite{fan2017computation}, $w$                       & $[70,130]$mW\\
\hline
\end{tabular}
\label{table:table1}
\end{table}

\begin{table}
\centering
\caption{Energy consumption comparison}
\begin{tabular}{|c| c|} 
 \hline
  \textbf{Semantic Data Transmission} & \textbf{Non-semantic Data Transmission} \\ 
 \hline
 0.896J                                             & 111J\\
\hline
\end{tabular}
\label{table:table3}
\end{table}

Besides the two decision variables listed in Section~\ref{dip}, there is one more decision variable in the SIP formulation. $m_{w,e}^{(\mathrm{o})}(\lambda_i)\in\{0,1,\ldots\}$ indicates the number of semantic data transmissions that VSP $w$ is requested on-demand from the edge device $e$ in scenario $\lambda_i$.

The objective function given in~(\ref{sip1}) and~(\ref{sip2}) is to minimize the cost of resource allocation. The expressions in~(\ref{sip1}) and~(\ref{sip2}) represent the first- and second-stage SIP, respectively. The SIP formulation can be expressed as follows:

\noindent $\displaystyle\min_{m_{w,e}^{(\mathrm{r})}, \bar{m}_{w,e}^{(\mathrm{r})},m_{w,e}^{(\mathrm{o})}(\lambda_i)}$:

 \begin{align}\label{sip1}
    \sum_{w\in \mathcal{W}}\sum_{e\in\mathcal{E}}\biggl(m_{w,e}^{(\mathrm{r})}C_e^{(r,memb)} + \bar{m}_{w,e}^{(\mathrm{r})}C_e^{(r,trans)} \biggr) +\nonumber\\ \mathbb{E}\biggl[\mathcal{Q}(m_{w,e}^{(\mathrm{o})}(\lambda_i)\biggr],
\end{align}
where
\begin{align}\label{sip2}
    \mathcal{Q}(m_{w,e}^{(\mathrm{o})}(\lambda_i)) =\sum_{\lambda_i\in\Omega}P(\lambda_i)\sum_{w\in \mathcal{W}}\sum_{e\in\mathcal{E}}
    m_{w,e}^{(\mathrm{o})}(\lambda_i)C_e^{(o,trans)},
\end{align}
subject to: (\ref{dip_cons1}), (\ref{dip_cons3})-(\ref{dip_cons4})
\begin{align}
    \sum_{e\in\mathcal{E}}\bar{m}_{w,e}^{(\mathrm{r})}nS_{w,e}(\lambda_i) + \sum_{e\in\mathcal{E}}m_{w,e}^{(\mathrm{o})}(\lambda_i) \geq \bar{F}_{w}(\lambda_i)\Tilde{F}_{w}(\lambda_i), \nonumber\\\forall w\in \mathcal{W},\forall\lambda_i\in\Omega,\label{sip_cons1}
\end{align}
\begin{align}
    m_{w,e}^{(\mathrm{o})}(\lambda_i)\in \mathbb{Z}\text{-}*, \hspace*{+11mm}\forall w\in\mathcal{W}, \forall e\in\mathcal{E}, \forall\lambda_i\in\Omega.\label{sip_cons2}
\end{align}

(\ref{sip_cons1}) is similar to \eqref{dip_cons2}, it is to ensure the demand is met. (\ref{sip_cons2}) indicates that $m_{w,e}^{(\mathrm{o})}(\lambda_i)$ is non-negative integer.

\begin{figure*}
\centering
\begin{multicols}{3}
\includegraphics[width=0.8\columnwidth]{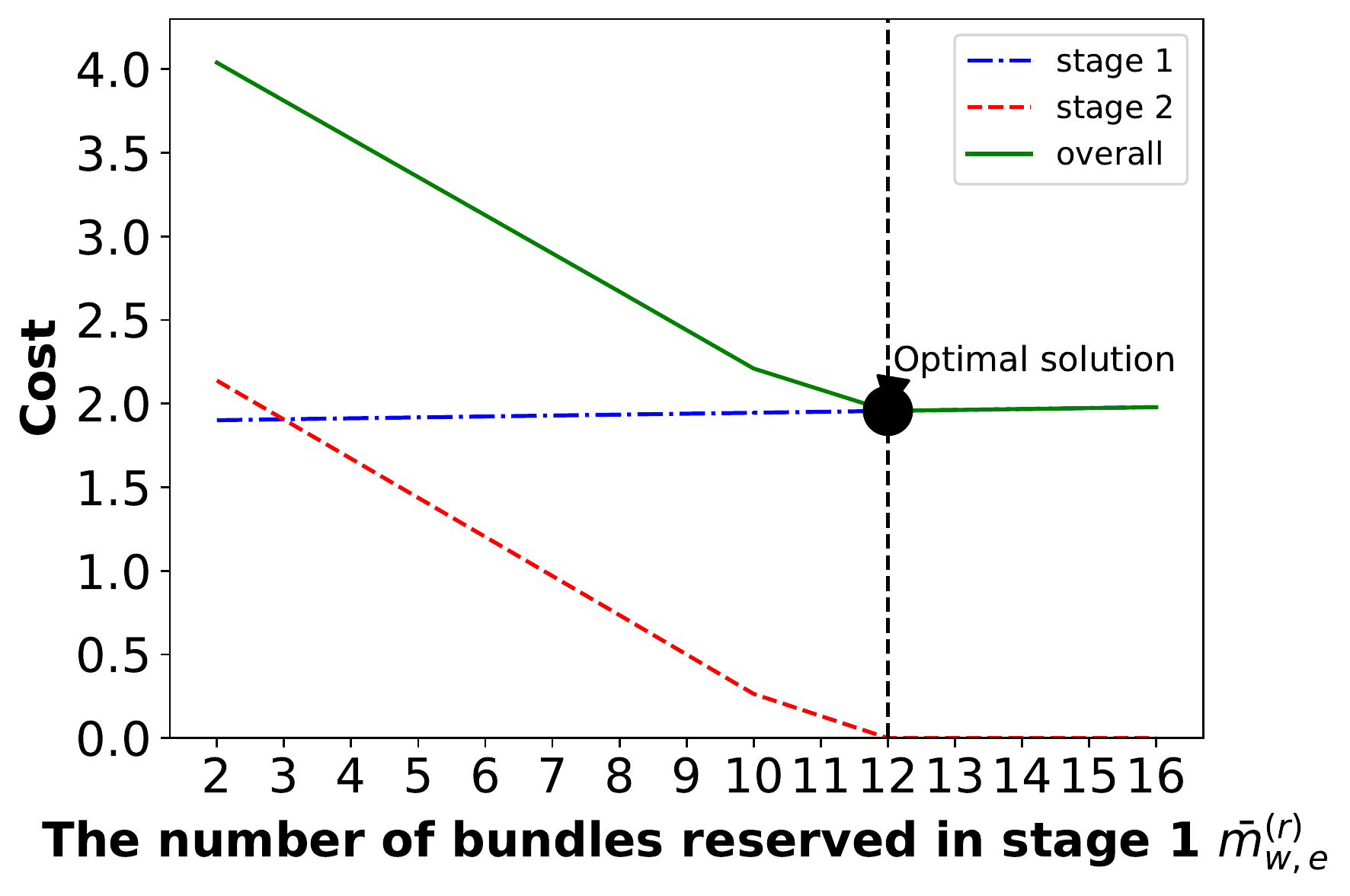}
\caption{The optimal solution in a simple SIP network.}
\label{fig:optimal cost}
\includegraphics[width=0.7\columnwidth]{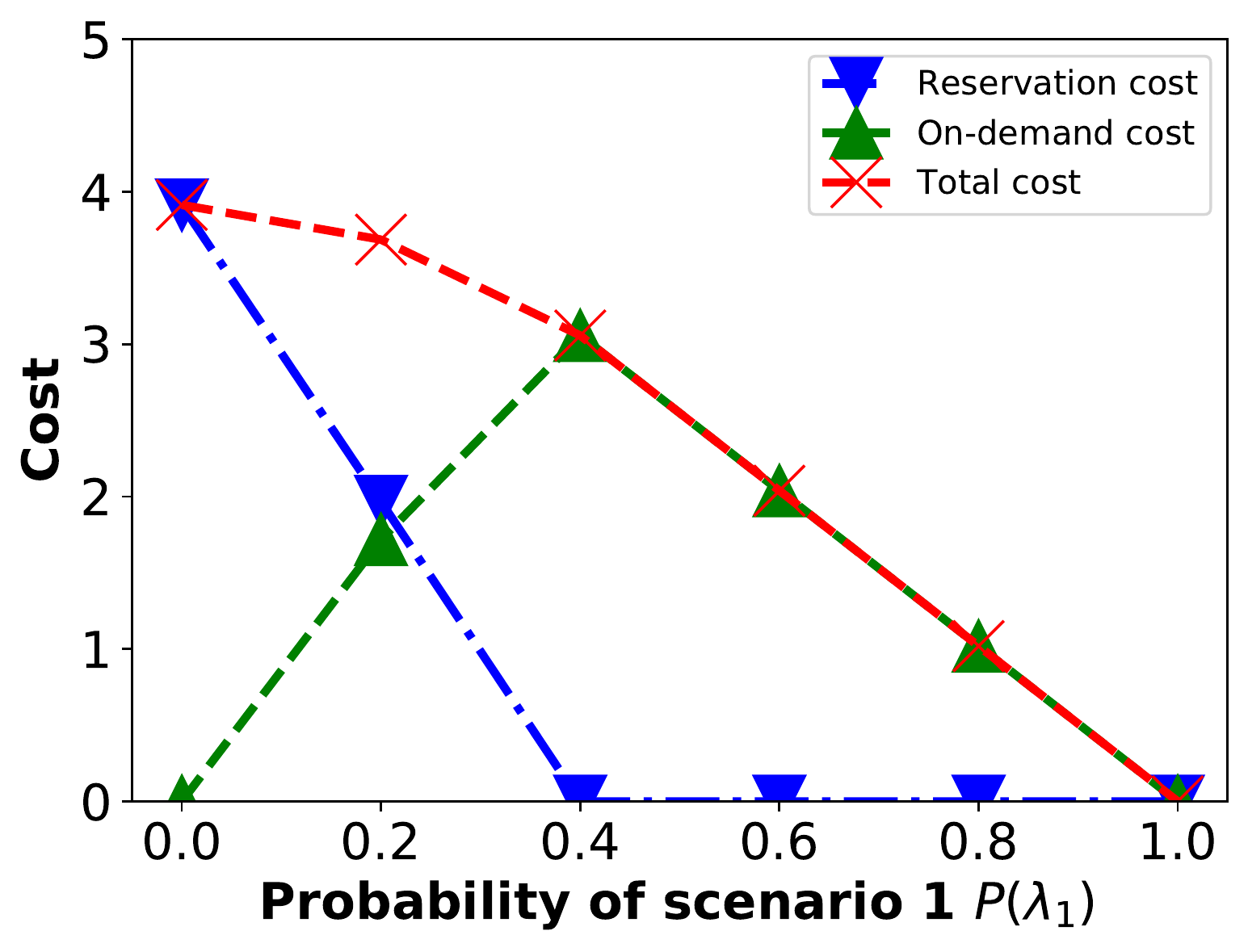}
\caption{The cost of the network when the probability of the scenarios are varied.}
\label{fig:probability}
 \includegraphics[width=0.8\columnwidth]{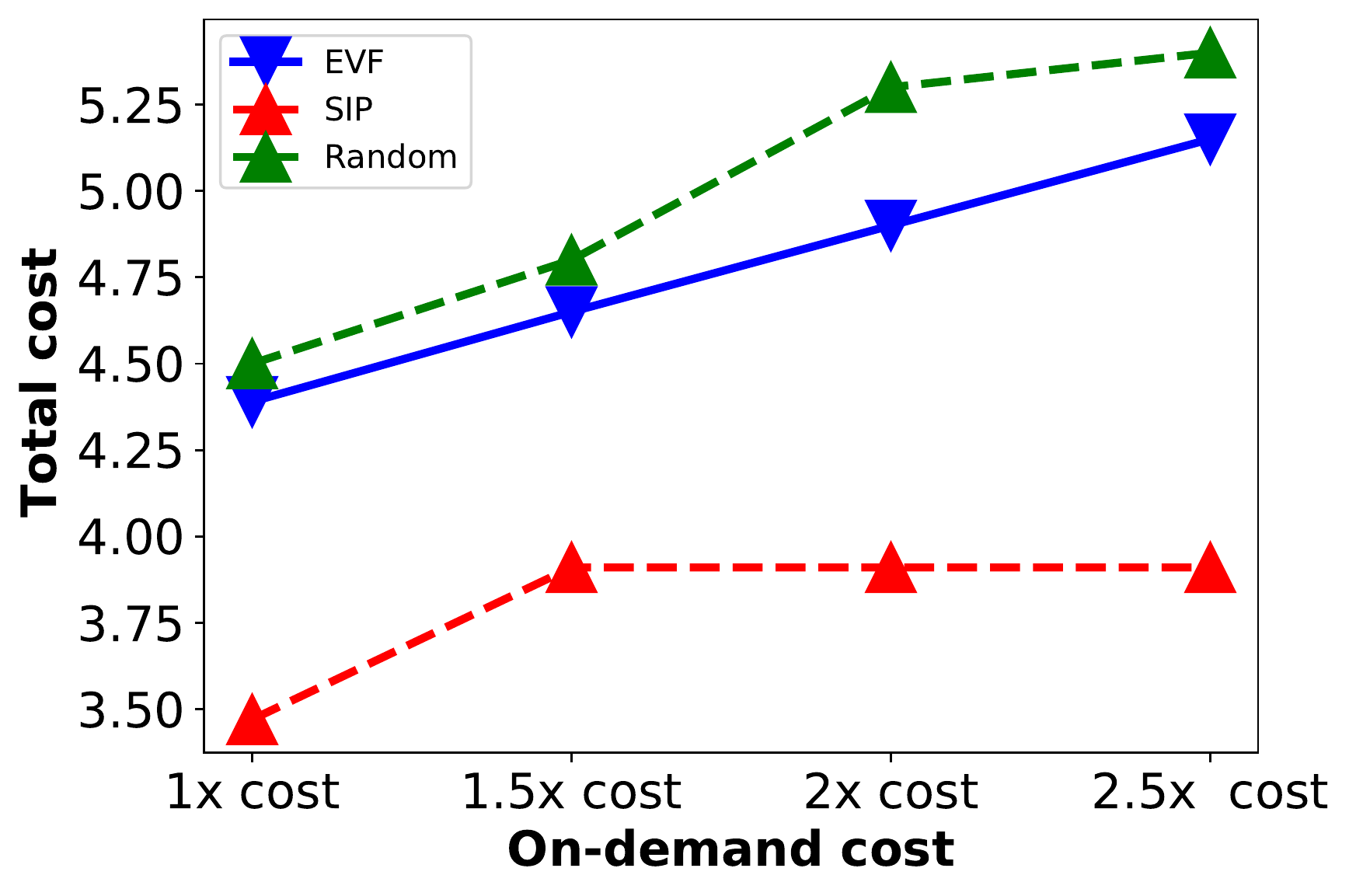}\par
  \caption{SIP comparing with EVF and random schemes.}
  \label{fig:evf}
\end{multicols}
\vspace*{-7mm}
\end{figure*}

\section{Performance Evaluation}\label{simulation}
For the simulations, SSTS initializes two VSPs and deploy three smartphones $e_1$, $e_2$, and $e_3$, around Singapore. The smartphones (model: iPhone 13 Pro Max) are represented by the yellow, blue and red markers respectively in Fig \ref{fig:case}. We consider the daily rental cost is the membership cost of each smartphone, where $C_e^{r,memb}=\$1.89$~\cite{cameracost}. $C_e^{r,trans}$ and $C_e^{o,trans}$ are additional cost for the transmission of semantic data. The simulation parameters are summarized in Table~\ref{table:table1}.

To solve the SIP, we consider that the probability distribution of all scenarios in set $\Omega$ is known~\cite{dyer2006computational}. For the presented experiments, we implement the SIP model using GAMS script~\cite{chattopadhyay1999application}.

\subsubsection{Energy efficiency} We first compare the energy consumption with and without semantic communication. An image's average transmitted data size is $5.2$Mb, while the average transmitted semantic data is $41$Kb. The reason is that the transmitted semantic data is precisely  the region that the VSP is interested in (demand). Using SSTS, we further perform the resource allocation and record the energy consumed with and without semantic communication. The result is shown in Table~\ref{table:table3}. One instance of semantic data only requires 0.896J of energy in the transmission, whereas non-semantic data requires 111J of energy. Semantic extraction (using YOLO) also requires very little energy to compute~\cite{kim2020spiking}. Therefore, with the help of semantic data, edge devices can reduce their power consumption during transmission as well as storage costs, which means that they will charge the VSPs less as the transmission cost depends on the transmission energy. In addition, it improves the sustainability of developments in the Metaverse.

\subsubsection{Cost structure} We then study the cost structure of the network. As an illustration, a primitive network is considered with $w_1$, $e_1$, and one demand scenario $|\Omega|=1$. VSP has a demand to require a certain amount of semantic data from the smartphone. We observe the cost structure of the network by varying the number of bundles reserved in the first stage $\bar{m}^{(r)}_{w,e}$. In Fig.~\ref{fig:optimal cost}, the costs in the first stage and second stage, and the total cost under the different number of the bundles reserved $\bar{m}^{(r)}_{w,e}$ are presented. We can observe that the first stage cost (reservation cost) increases as the number of bundles reserved increases. With more bundles reserved in the first stage, the cost in stage 2 is reduced as the need for on-demand transmissions is less likely. It can be identified that even in this simple network, the optimal solution is not trivial to obtain due to the uncertainty of demands. For example, the optimal cost is not the point where the costs in the first and second stages intersect. Therefore, the SIP formulation is required to guarantee the minimum cost to the network.

\subsubsection{Probability of demand scenario (has interest or no interest)}\label{probability_demand} Next, we consider two demand scenarios $|\Omega|=2$. In the first scenario $\lambda_1$, both VSPs 1 and 2 have no demand. In the second scenario $\lambda_2$, both VSPs 1 and 2 have demands. VSP 1 requires $200$ semantic data transmissions while VSP 2 requires $300$ semantic data transmissions. We analyze the first stage (reservation), the second stage (on-demand), and the total cost by varying both the demand probabilities $P(\lambda_1)$ and $P(\lambda_2)$. Since $P(\lambda_1) + P(\lambda_2) =1$, $P(\lambda_2)=1$ when $P(\lambda_1)=0$. Figure~\ref{fig:probability} depicts the network cost. When $P(\lambda_1)=0$, both VSPs 1 and 2 subscribe to the reservation plan and pay the corresponding subscription fee as they will always have demands, and it is definitely cheaper to use the reservation plan than the on-demand plan. When $P(\lambda_1)$ increases to 0.2, VSP 1 changes its decision and uses the on-demand plan instead of the reservation one. VSP 2 continues to subscribe using the reservation plan. This is due to the fact that $P(\lambda_2)$ decreases as $P(\lambda_1)$ increases, and when there is a demand, VSP 1 only requires 200 semantic data transmissions. It is less than VSP 2, which requires 300 semantic data transmissions. Moreover, there is an additional membership fee in the reservation plan. Therefore, It is cheaper for VSP 1 to subscribe to the on-demand plan only when demand occurs. When $P(\lambda_1)\geq0.4$, the on-demand plan is cheaper than the reservation plan for both VSPs 1 and 2, and the total cost reduces as $P(\lambda_1)$ increases. Eventually, the total cost is zero when $P(\lambda_1)=1$ as both VSPs have no demand.

\begin{table}
\centering
\caption{Varying the demand interest of VSP 1}
\small
\begin{tabular}{|c|c|c|c|c|c|c|} 
 \hline
  \multirow{2}{*}{Variables} & \multicolumn{6}{c|}{$P(\lambda_1)$}\\\cline{2-7}
                                 & 0 &0.2 &0.4 &0.6 &0.8 &1\\
 \hline
 $\bar{m}^{(r)}_{1,1}$   & 11 &12 &12 &12 &12 &0\\
 $\bar{m}^{(r)}_{1,3}$   & 0 & 0 & 0 & 0 &0 &10\\
 $\bar{m}^{(o)}_{1,3}(\lambda_1)$  &2 &0 &0 &0 &0 &0\\
\hline
\end{tabular}
\label{table:table2}
\end{table}
\subsubsection{Probability of demand scenario (different interest)} Different from the setup in Section~\ref{probability_demand}, we study the VSP's decision when its interest varies. For ease of exposition, we only consider a single VSP under two demand scenarios $|\Omega|=2$. In the first scenario $\lambda_1$, VSP 1 has an interest in vehicles on road. In the second scenario $\lambda_2$, VSP 1 has an interest in buses and traffic lights. The average similarity scores for smartphones 1, 2, and 3 in scenario 1 are $0.72$, $0.697$, and $0.83$, respectively. The average similarity scores for smartphones 1, 2, and 3 in scenario 2 are $0.793$, $0.661$, and $0.57$, respectively. The simulation result is shown in Table~\ref{table:table2}. Due to a large number of variables, the table only shows the variables of value. When $P(\lambda_1)=0$, VSP 1 purchases 11 bundles from smartphone 1 using the reservation plan and 2 additional images using the on-demand plan as smartphone 1 has the highest average similarity score. When the probability increases, VSP 1 purchases 12 bundles from smartphone 1 by using the reservation plan. The additional bundle is used to balance the shortfall from scenario 1. When $P(\lambda_1)=1$, the demand is met by using only 10 bundles from smartphone 3, which has a higher similarity score.

\subsubsection{Comparing between EVF, SIP and random scheme} We compare the SIP with other baselines such as expected-value formulation (EVF)~\cite{ng2022unified} and random scheme. EVF is an approximation scheme that uses the average demand to solve the DIP and uses the solution as the first stage decision variable value in SIP. In the random scheme, the values of the decision variables are randomly generated. We vary the on-demand cost to compare the difference between EVF, SIP, and random schemes. Figure.~\ref{fig:evf} depicts the comparison result. As shown in the result, the EVF and random schemes cannot adapt to the change in cost. Unlike the SIP scheme, when the VSPs realize that the on-demand cost is high, the VSPs change their subscription plan to on-demand, and this is the reason why the total cost for the SIP scheme remains constant when the on-demand cost is $\geq 1.5\times$

\section{Conclusion}\label{conclusion}
In this paper, we have presented a resource allocation framework, SSTS for the Metaverse, in a case study of utilizing semantic communication to develop the virtual transportation network. To achieve the optimal allocation, SSTS minimizes the total cost of the network even amid demand uncertainty. Using real data, our numerical studies and simulations have validated that SSTS reduces transmission costs and achieves the best solution as it can better adapt to changes in the probability of users' demands.

\bibliographystyle{IEEEtran}
\bibliography{mybibliography.bib}
\end{document}